\documentclass[preprint,showpacs,preprintnumbers,amsmath,amssymb]{revtex4}
\usepackage{graphicx}
\begin{document}

\title{Prospects for GMRT to Observe Radio Waves from UHE Particles Interacting with the Moon}

\author{Sukanta Panda$^{a}$, Subhendra Mohanty$^{b}$, Padmanabhan Janardhan$^{c}$, and Oscar St\aa l$^d$}%
\affiliation{$^a$  Department of Physics, Universitat
Aut{\`o}noma de Madrid, Madrid, Spain\\
$^b$ Physical Research Laboratory, Ahmedabad - 380 009, India.\\
$^c$ Instituto Nacional de Pesquisas Espaciais (INPE); Divisao de Astrofisica; Brazil.\footnote{On Sabbatical leave from the
Physical Research Laboratory, Ahmedabad - 380 009, India.}\\
$^d$ Department of Nuclear and Particle Physics, Uppsala University, P.\,O.\,Box 535, SE-751 21 Uppsala, Sweden}
\def\be{\begin{equation}}
\def\ee{\end{equation}}
\def\al{\alpha}
\def\bea{\begin{eqnarray}}
\def\eea{\end{eqnarray}}

\begin{abstract}
Ultra high energy (UHE) particles of cosmic origin impact the lunar regolith and produce radio signals through Askaryan effect, signals that can be detected by Earth based radio telescopes. We calculate the expected sensitivity for observation of such events at the Giant Metrewave Radio Telescope (GMRT), both for UHE cosmic rays (CR) and UHE neutrino interactions. We find that for 30 days of observation time a significant number of detectable events is expected above $10^{20}$ eV for UHECR or neutrino fluxes close to the current limits. Null detection over a period of 30 days will lower the experimental bounds by a magnitude competitive to both present and future experiments at the very highest energies.
\end{abstract}

\maketitle

\section{Introduction}
The study of cosmic rays with energies near and above the Greisen-Zatsepin-Kuzmin (GZK) cut-off $\sim5 \times 10^{19}$ eV \cite{gre,zat} is of great interest in particle astrophysics. Above the GZK energy, proton interactions $p+\gamma_\mathrm{CMB}\to N\pi$ on Cosmic Microwave Background (CMB) photons are possible. The universe in effect becomes opaque over Mpc scales. To check the existence of this cut-off is therefore an important motivation for cosmic ray experiments like AUGER \cite{auger} and HIRES \cite{hires}.
Complementing the cosmic ray experiments, a number of experiments perform direct searches for ultra high energy (UHE) neutrinos from the GZK pion decays ($\pi^+\to\mu^+\nu_\mu\to e^+\nu_e\bar{\nu}_\mu\nu_\mu$). The search for neutrinos above the GZK energy is also motivated in the context of Grand Unification Theories (GUT) \cite{TD}. Relic 'X'-particles with ultra high mass $M_X\sim10^{22}$~eV originating from the GUT phase-transition in the early universe could be decaying in the present epoch. Thereby they would produce UHE neutrinos. As a final example of a theoretically very appealing possibility, is the Z-burst process \cite{Zb1}. In this scenario, the highest energy cosmic rays are produced following resonant interactions ($\nu\bar{\nu}\to Z^0$) of UHE neutrinos on the relic neutrino background. To explain the full cosmic ray spectrum, this would require a very high flux of neutrinos in the $E_\nu\sim10^{22}$ regime to be present. Limits constraining the possible UHE !
 neutrino flux from these processes exist, with the most stringent one to date being presented by the ANITA-lite experiment \cite{anita}.

Among the various methods for detecting ultra high energy (UHE) particles,
a promising technique utilizes the Askaryan
effect \cite{aska}. This effect causes electromagnetic cascades, induced by the UHE particle interactions in dense media, to develop a negative charge excess. The charge excess will radiate coherently at radio wavelengths, thereby producing strong coherent pulses of \v{C}erenkov radiation. In a series of accelerator experiments, the Askaryan
mechanism has been confirmed to work in different media. Most important for our purposes, it was tested in Silica sand specifically to mimic the conditions of the lunar material \cite{slac1}. It was first proposed by Askaryan \cite{aska}, later by Dagkesamanskii and
Zheleznyk \cite{dz}, that the radio transparent lunar regolith
(the 10 - 20 m deep surface layer of the Moon,
consisting mainly of fractured rock) would be an ideal target for studying
such UHE particles. An additional advantage is the absence of an atmosphere which implies that electromagnetic showers in the lunar regolith are caused by the primary cosmic particles.

A number of attempts at observing
cosmic ray induced radio waves from the Moon have since been carried out using
ground based radio telescopes. The first experiment \cite{parkes} used
the 64 m diameter Parkes radio telescope in Australia to make coincidence
measurements from two polarization channels. A 500-MHz band centered at 1.425 GHz was used after making appropriate corrections for the ionospheric delay between two sub-bands.
The second was the Goldstone Lunar Ultra-high energy
neutrino Experiment (GLUE) \cite{glue} that used two large telescopes of the JPL/NASA Deep Space Network in Goldstone (CA). A third experiment of this kind
took place at Kalyazin Radio Astronomical Observatory, using again a single 64 m radio telescope \cite{kalyazin}. None of these experiments however, resulted in any detection of signals from UHE particles. Under the name NuMoon, there is one currently on-going search using the Westerbrok Synthesis Radio Telescope (WSRT) in the Netherlands \cite{Scholten}.
For future experiments with increased sensitivity there are proposals \cite{aska3,Falcke,Scholten:2005pp} to study
this effect with upcoming radio telescopes like the Low Frequency Array (LOFAR) and the Square
Kilometer Array (SKA), or to use lunar orbiting artificial satellites carrying one or more radio antennas \cite{gusev,Stal}.

In this paper we study in detail the possibility of detection
of UHECR and neutrino radio waves from the Moon with the
Giant Metrewave Radio Telescope (GMRT)
facility \cite{gmrt-web}. The GMRT comprises $30$ fully
steerable dishes spread over distances up to $25$ km, each with a diameter of $45$ m.
Half of these antennas are distributed randomly over about $1$ km$^2$ in a
compact array while the rest are spread out in an approximate "Y"
configuration. The GMRT currently operates in eight frequency
bands around $150$, $235$, $325$, $610$ and
$1000$-$1450$ MHz. The shortest baseline
is $100$ m while the longest one is $26$ km. The telescope has an effective area in total of $30000$ m$^2$
for frequencies up to $327$ MHz and $18000$ m$^2$ at the higher frequencies. The RMS sensitivity of
the GMRT, using a $2$ s integration time and a bandwidth of $32$ MHz is $0.3$ mJy and $0.03$ mJy at $327$ MHz and
$1.4$ GHz respectively. With such high sensitivity, we expect a substantial improvement from previous experiments in the
measurement of the Askaryan radio waves from the Moon.

\section{Radio waves from cascades in the lunar regolith}
High energy charged particles from UHE cosmic ray (CR) or UHE neutrino interactions in the lunar regolith will initiate a cascading shower with total energy $E_s$ and typical length scale \cite{Scholten:2005pp}
\begin{equation}
L(E_s) = 12.7 + \frac{2}{3}\log_{10}\frac{E_s}{10^{20}\,\mathrm{eV}}
\end{equation}
in units of radiation length. For the lunar regolith with radiation length
 $X_0=22.1$ g/cm$^2$ and density $\rho=1.7$ g/cm$^{-3}$, the shower length for a particle with energy $E=10^{20}$ eV is therefore $1.7$ m. Since the particles in the cascade travel at speeds very close to the speed of light the duration of the shower is $L/c\simeq5.67$\,ns. The radio waves of
about $6$ ns duration which are emitted from the lunar
regolith are dispersed as they propagate through
the atmosphere of the Earth. The electrons in the
ionosphere cause a time delay \cite{Falcke} of
\begin{equation}
t_d(\nu)=1.34\times10^{-7}\left(\frac{N_e}{\mathrm{m}^{-2}}\right) \left(\frac{\mathrm{Hz}}{\nu}\right)^2
\end{equation}
seconds, where $N_e=10^{17}$ m$^{-2}$ is the typical night-time column density of electrons. The dispersion of a radio signal of frequency $\nu$ and bandwidth $\Delta \nu$ at zenith is therefore
\begin{equation}
\Delta t_d =134 \, \mathrm{ns}\times\, \left(\frac{\Delta \nu}{40\, \mathrm{MHz}}\right) \left(\frac{200\, \mathrm{MHz}}{\nu} \right)^3.
\end{equation}
One must thus look for \v{C}erenkov radio pulses of about $100$ ns duration from the Moon and use coherent de-dispersion to recover the broadband structure of the signals.

\begin{figure}[t]
\hbox{\hspace{0cm}
\hbox{\includegraphics[scale=1.3]{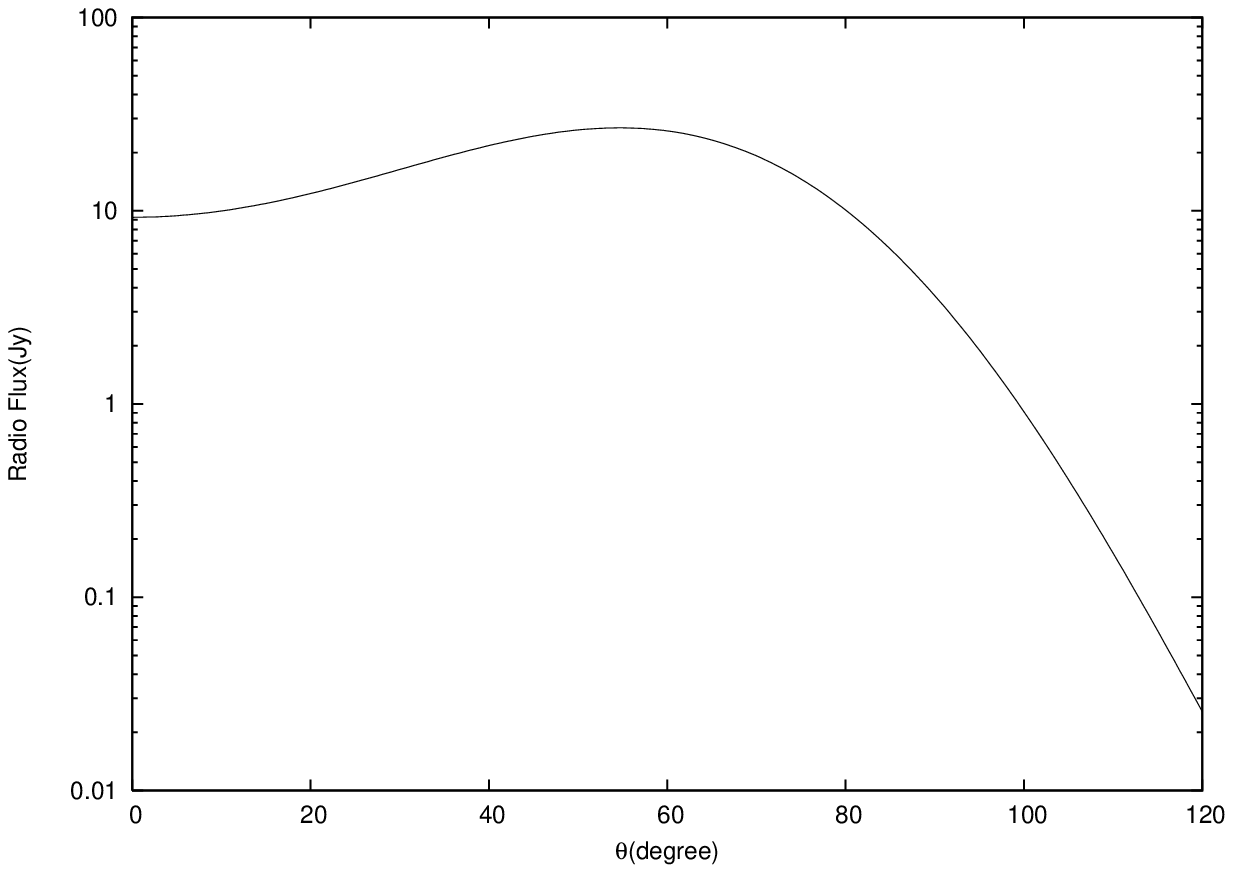}}}
\caption{Radio flux at Earth from the center of the Moon vs.
emission angle $\theta$ at $\nu=150$ MHz and bandwidth $\Delta\nu=40$ MHz
for a shower initiated with $E_s=10^{20}$ eV. The dashed line contains the complete $\theta$-dependence in Eq.~(\ref{f}), whereas the full line shows the result of using the Gaussian approximation.}
\label{radioflux1}
\end{figure}

The intensity of radio waves on Earth from a shower in the lunar regolith with energy $E_s$ emitting \v{C}erenkov emission has been parameterized from simulations \cite{ZHS,Scholten:2005pp}. At an angle $\theta$ to the shower axis, for radiation frequency $\nu$ and bandwidth $\Delta \nu$, it is given by
\begin{equation}
F(E_s,\theta,\nu)= 5.3 \times 10^{5} \, f(E_s,\theta) \left(\frac{E_s}{\mathrm{TeV}}\right)^2 \times
\left(\frac{1\,\mathrm{m}}{R} \right)^2 \left(\frac{\nu}{\nu_0[1+(\nu/\nu_0)^{1.44}]}\right)^2 \, \frac{\Delta \nu}{100\,\mathrm{MHz}}\,\,\, \mathrm{Jy}
\label{F}
\end{equation}
where $R$ is the distance between the emission
point on the Moon's surface to the telescope and $\nu_0=2.5$ GHz.
Furthermore there is an angular dependence given by
\begin{equation}
 f(E_s,\theta)=\left(\frac{\sin\theta}{\sin\theta_C}\right)^2\exp(-2Z^2)
\label{f}
\end{equation}
 with $Z= \sqrt{\kappa} (\cos\theta_C-\cos\theta)$ and $\kappa(E_s)=(\nu/\mathrm{GHz})^2(70.4 + 3.4 \ln(E_s/\mathrm{TeV}))$. The angular distribution of the radiation pattern is illustrated in Fig.~\ref{radioflux1}. In this figure we also show the commonly used Gaussian approximation where the forward-suppression factor $\sin^2\theta$ in (\ref{f}) is ignored. For high frequencies this has no effect. For low frequencies, the differences at small angles only plays a role for showers nearly parallel to the surface normal, while the effects of changing the normalization near the \v{C}erenkov angle is important also for more horizontal showers. A measure of the effective angular spread $\Delta_C$ of the emission around the \v{C}erenkov angle $\theta_C$ is given in terms of
\begin{equation}
\Delta_C=\sqrt{\frac{\ln(2)}{\kappa(E_s)}}\frac{1}{\sin\theta_C}.
\end{equation}
For the example in Fig.~\ref{radioflux1} this becomes $\Delta_C\simeq44^\circ$.
 The \v{C}erenkov radio waves are expected to be $100 \%$ linearly polarized \cite{ZHS}. Radio signals should therefore be observed with both the LCP and RCP modes. A simultaneous triggering of both the LCP and RCP modes (with $50 \%$ of the total intensity in each mode) will be a good signature of the transient \v{C}erenkov radio wave emission from UHE particles. In addition, the large number of GMRT dishes should allow pointing discrimination and spatial-temporal coincidence measurements to be used for signal identification and background rejection.

\section{Aperture for cosmic ray and neutrino events at GMRT}
For cosmic rays, which produce showers by
hadronic processes, the shower energy $E_s$ is equal to
the energy of the primary particle. When instead
the showers are produced by deep inelastic neutrino scattering off target nucleons, the fraction of the total energy which produces the hadronic showers
is, on average, $E_s=0.25 E_P$ \cite{Gandhi}.

We model the lunar regolith as a homogeneous medium with dielectric
permittivity $\epsilon=3$, hence refractive index $n=1.73$. The \v{C}erenkov angle is  $\theta_C= 54.7^\circ$.  The radio waves are also absorbed in the regolith, which we take to have a frequency dependent attenuation length $\lambda_r= 9/(\nu/\mathrm{GHz})$ m \cite{lunar-source}.
This means the attenuation length of radio waves is always much smaller than the
mean free path of the neutrinos, conveniently parameterized as $\lambda_\nu=670(\mathrm{1\,EeV}/E_{\nu})^{0.363}$ km using the cross sections of \cite{Gandhi}. As can be seen from this expression, the mean free path for an UHE neutrino is not sufficient to escape the Moon. Most of the passing neutrinos will therefore interact in the material.
\begin{figure}[t]
\hbox{\hspace{0cm}
\hbox{\includegraphics[scale=1]{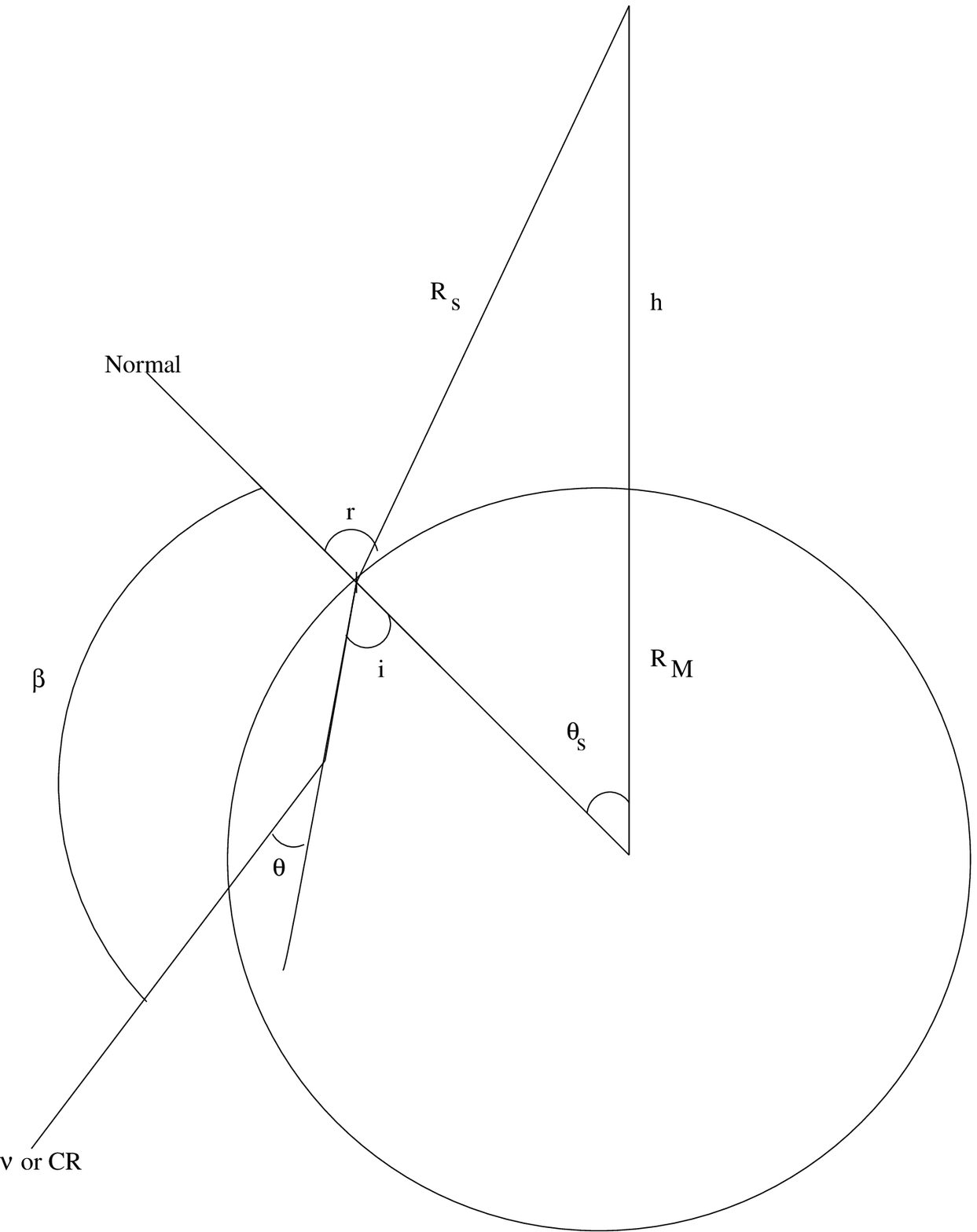}}}
\caption{Geometry of an UHE CR or neutrino event which generates \v{C}erenkov radiation of radio waves in the lunar regolith. The Earth is located at a mean distance $h$ from the lunar surface.}
\label{geometry}
\end{figure}

 The threshold energy required for showers to be detectable is determined by the sensitivity of the radio telescope. This is commonly presented in terms of a noise intensity
\begin{equation}
F_N= \frac{2 k_B T_{sys}}{\sqrt{\delta t \Delta \nu} A_\mathrm{eff}},
\end{equation}
where $T_\mathrm{sys}$ is the system temperature, $k_B$ is Boltzmann's constant,
$\delta t$ the integration time, $\Delta \nu $ the bandwidth, and $A_\mathrm{eff}$ the effective area of the telescope. In units of Jansky ($1$ Jy = $10^{-26}$ W/m$^2$/Hz) the noise intensity is given by
\begin{equation}
F_N= 2.76 \times 10^3  \frac{(T_\mathrm{sys}/\mathrm{K})}{\sqrt{\delta t
\Delta \nu} (A_\mathrm{eff}/\mathrm{m}^2)}.
\end{equation}
At GMRT, for frequencies $150, 235$ and $327$ MHz we use the effective area $A_\mathrm{eff}=30000$ m$^2$ while for the higher frequencies $A_\mathrm{eff}=18000$ m$^2$. The integration time $\delta t=1/\Delta_{\nu}$ for the bandwidth limited case. In Table 1 we list the system temperatures at the different observation
frequencies and the corresponding noise levels. Using equation (\ref{F}), we can solve for $E_s$ at the threshold required for measurement with the radio telescope (obtained for $\theta=\theta_C$ and $F=F_N$). If we take a required signal-to-noise ratio $\sigma$, the threshold shower energies $E_\mathrm{th}$ which can be measured at GMRT at the different observation frequencies are listed in Table 1.

\begin{figure}[t]
\hbox{\hspace{0cm}
\hbox{\includegraphics[scale=1.3]{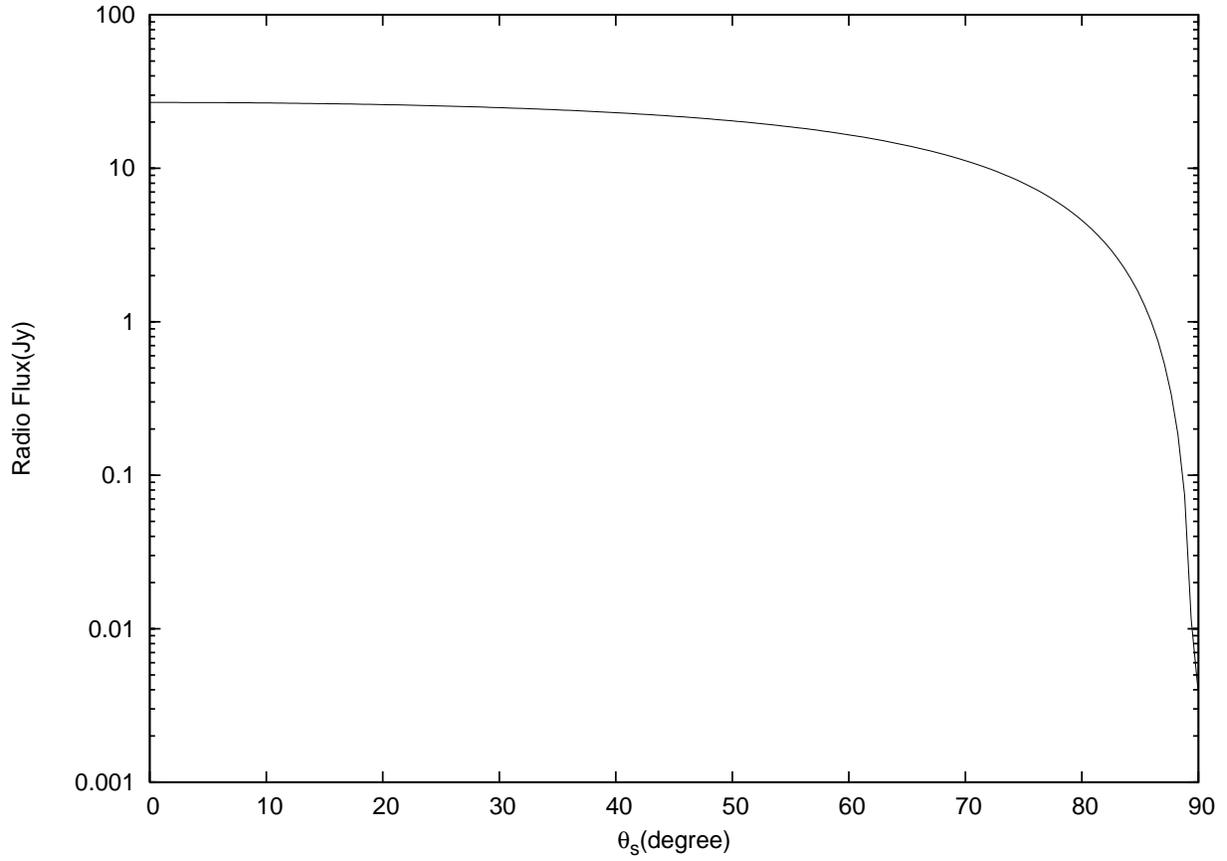}}}
\caption{Radio flux at Earth vs. the lunar polar angle
$\theta_s$ for $\theta=\theta_c$ at $\nu=150$ MHz and bandwidth $\Delta\nu=40$ MHz
for a shower initiated with $E_s=10^{20}$ eV.}
\label{flux_theta}
\end{figure}

The radio waves produced below the lunar surface get attenuated while propagating to the surface and they get refracted at the surface before reaching the telescope. The transmission coefficients are different for the electric field polarization which is parallel and perpendicular to the lunar surface. It is therefore convenient to express the \v{C}erenkov radiation intensity (\ref{F}) in terms of the electric field magnitude $\cal E$ of the radio waves. The conversion is
\begin{equation}
\frac{F}{\mathrm{Jy}}= \frac{10^{7}}{24 \pi} \left(\frac{{{\cal E}}}{\mu \mathrm{V}/\mathrm{m}/ \mathrm{MHz}}\right)^2\, \frac {\Delta \nu}{\mathrm{MHz}}
\end{equation}
Using this conversion we also list the threshold electric fields measurable at GMRT in Table 1. If attenuation is factored in separately, the quantity $R {\cal E}$ remains a constant between the point of emission and the point of refraction at the surface:
\begin{equation}
R {\cal E}= 0.2 \, e^{-Z^2} \left(\frac{E_s}{\mathrm{TeV}}\right) \times \left(\frac{\nu}{\nu_0[1+(\nu/\nu_0)^{1.44}]}\right)\, \frac{\mu \mathrm{V}}{\mathrm{MHz}} .
\label{RE}
\end{equation}
At the surface, there is partial reflection which changes the intensity of the outgoing electric field. The electric field observed at the telescope is determined relates to the incident electric fields by the Fresnel relations
\begin{eqnarray}
T_{\perp}=\frac{R {\cal E}_{\perp}}{(R {\cal E}_{\perp})_\mathrm{inc}}= \frac{2 \cos r}{n \cos i+ \cos r} \nonumber\\[8pt]
T_{\parallel}=\frac{R {\cal E}_{\parallel}}{(R {\cal E}_{\parallel})_\mathrm{inc}}=\frac{2 \cos r}{n \cos r+ \cos i}
\label{fresnel}
\end{eqnarray}
where the angles of incidence $i$ and refraction $r$ are related by Snell's law. The refraction angle is fixed by the polar angle $\theta_s$ through the geometrical condition
\begin{equation}
\sin(r)= \frac{\sin(\theta_s)(R_M+h)}{R(\theta_s)}.
\end{equation}
Here $R(\theta_s)=[R_M^2 + (R_M + h)^2 - 2 (R_M + h) R_M
\cos\theta_s]^{1/2}$ is the distance between the detector and the point of refraction on the lunar surface, and
$h=384\,000$ km is the average Earth-Moon distance. Putting everything together, the parallel component of the electric field at the telescope on Earth is given by
\begin{eqnarray}
{\cal E}_\parallel= \frac{1}{R(\theta_s)} \,
T_\parallel(\theta_s) ({R \cal E})_\mathrm{inc},
\label{Eperp}
\end{eqnarray}
where $ ({R \cal E})_\mathrm{inc}$ is given by (\ref{RE}) as a function of $\theta$ and $E_s$. The dependence on the polar angle $\theta_s$ of the radio flux on Earth is shown in Fig.~\ref{flux_theta}.

\begin{figure}[t]
\hbox{\hspace{0cm}
\hbox{\includegraphics[scale=1.3]{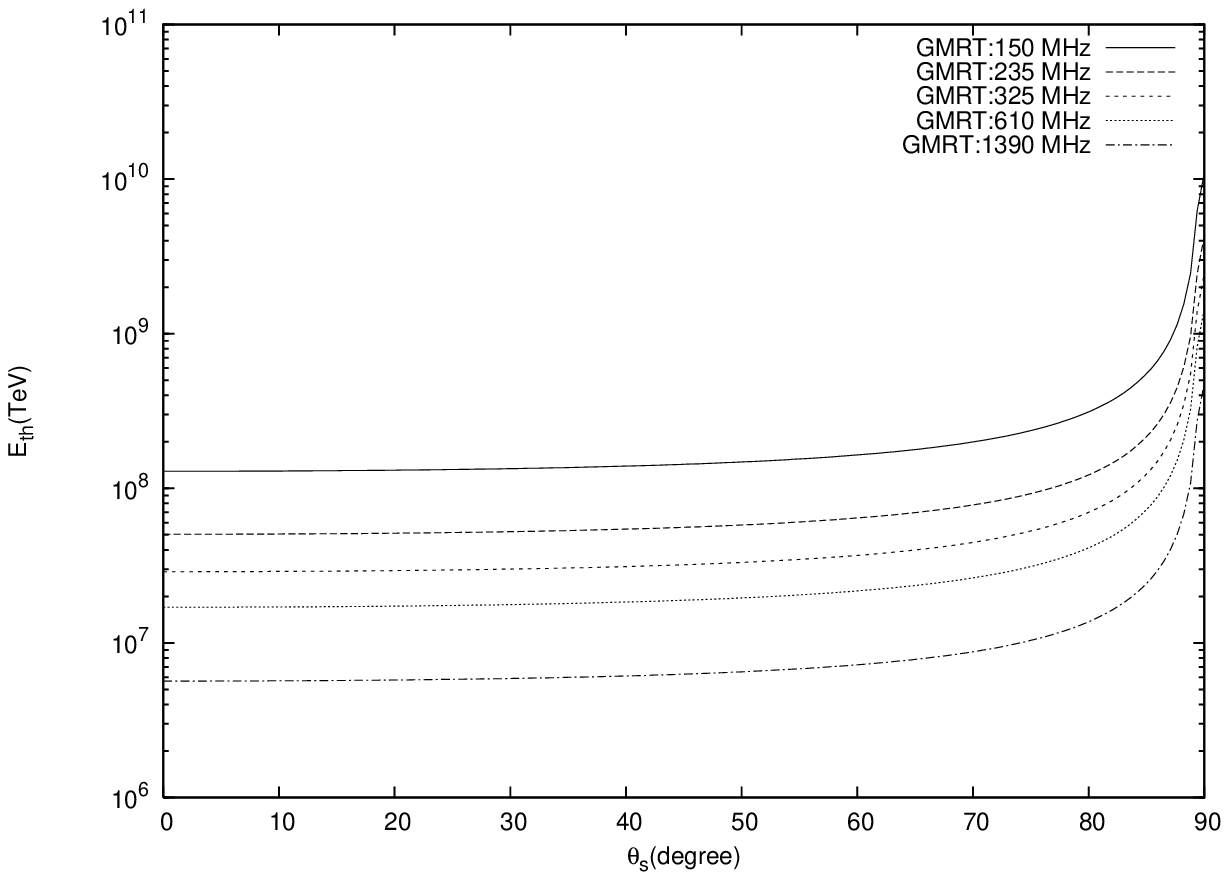}}}
\caption{Shower threshold energy vs. polar angle $\theta_s$ on the Moon for $\sigma=25$.}
\label{threshold}
\end{figure}

Evaluating  $ ({R \cal E})_\mathrm{inc}$ at
$\theta=\theta_C$ and equating ${\cal E}_\parallel$
with the threshold field ${{\cal E}}_\mathrm{th}$ which can be
measured with the telescope, we obtain the threshold
energy $E_{th}$ as a function  of the polar angle
$\theta_s$. Fig.~\ref{threshold} illustrates
that for observing the radio waves from the limb
of the Moon ($\theta_s \rightarrow \pi/2$ for $h\gg R_M$), one has to have a higher threshold energy
of the shower owing to the fact that the transmission
coefficients $T_\perp$ and $T_\parallel$ vanish
when $r \rightarrow \pi/2$. When $h$ is not large compared
to $R_M$ (which is the case for a telescope
aboard a satellite orbiting the Moon) then only a
small disc around the center of the Moon will be visible.

By equating the signal
${\cal E}_\parallel = \sqrt{\sigma} {\cal E}_{N}$ at the
\v{C}erenkov angle we determine the threshold $E_\mathrm{th}(\theta_s=0)$ at
different frequencies and bandwidths. The values so obtained are listed in Table~\ref{tab1}.

\begin{table}[htp]
\begin{tabular}{|l|l|l|l|l|l|l|r|}
\hline
$\nu$ (MHz) & $T_\mathrm{sys}$ (K) & $\Delta \nu$ (MHz) & $F_N$ (Jy) &
${\cal E}_N \,$($\mu$V/m/MHz) & $E_\mathrm{th}(0)/\sqrt{\sigma}$ (eV)
\\ \hline
  150 & 482 & 40 & 44 & .0029 & $1.28 \times 10^{20}$ \\
  235 & 177 & 40 & 16 & .0017 & $5.04 \times 10^{19}$ \\
  325 & 108 & 40 & 10 & .0013 & $2.9 \times 10^{19}$ \\
  610 & 101 & 60 & 15 & .0014 & $ 1.68 \times 10^{19}$ \\
  1390& 72 & 120 & 11 & .0008 & $ 5.6 \times 10^{18}$ \\
\hline
\end{tabular}
\caption{GMRT parameters, sensitivity and threshold energy. ${F}_N$ is the noise intensity of GMRT and $\cal{E}_N$ the corresponding electric field. The threshold energy in the last column is calculated with $\sigma=25$.}
\label{tab1}
\end{table}

The event rate that would be expected at the telescope can be related to an isotropic flux $\Phi$ of UHE particles on the Moon through
\begin{equation}
\frac{\mathrm{d}N_i}{\mathrm{d}t}= \int \frac{\mathrm{d}\Phi_i}{\mathrm{d}E} A_i(E) \mathrm{d}E,
\end{equation}
where $i=\{\mathrm{CR},\nu\}$ denote the type of primary particle and $A_i(E)$ is an aperture function corresponding to the effective detector area. We will differentiate between neutrinos, which have a large mean free path and therefore can penetrate the interior of the Moon before producing the \v{C}erenkov radio waves, and the strongly interacting cosmic rays (referred to by the subscript $\mathrm{CR}$) which can penetrate only small distances below the surface of the Moon. The aperture can be further decomposed into an angular aperture $\Delta\Omega_i(E,\theta_s)$ and a geometric area factor for the Moon
\begin{equation}
A_{i}(E)= 2 \pi R_{M}^2 \int \Delta\Omega_{i}(E,\theta_s)
\mathrm{d}\cos\theta_s
\label{acr}
\end{equation}
with $R_M=1760$ km. To evaluate the aperture, we use the analytical methods described in \cite{gusev}. For the case of cosmic rays, the angular aperture is given by
\begin{equation}
\Delta\Omega_{CR}(E,\theta_s)=\int \cos\beta
\Theta[{\cal E} (E,\theta_s)-{\cal E}_{th}] \times \Theta(\cos\beta)
\mathrm{d}\alpha\mathrm{d}\cos\beta,
\label{angle}
\end{equation}
where $\beta$ and $\alpha$ are the polar and azimuthal coordinates of the ray normal to the Moon's surface in a system where the shower direction defines the $z$ axis. The full geometry and the different angles are described in Fig.~\ref{geometry}.
Using the Gaussian approximation for the radiation pattern, the angular aperture for cosmic rays is obtained by directly integrating over the shower coordinates $(\theta,\phi)$:
\begin{equation}
\Delta \Omega_{CR}= 2 \int_{1/n - D}^{\sin(i)} \mathrm{d}\cos\theta
 \int_{0}^{\phi_m} \mathrm{d}\phi[\sin(\theta) \sin(i) \cos(\phi) - \cos(\theta) \cos(i)].
\end{equation}
Here $\cos(\phi_m)=\frac{\cos(i) \cos(\theta)}{\sin(i) \sin(\theta)}$, and in the lower integration limit $D=\sqrt{\frac{1}{\kappa} \ln\frac{E}{E_\mathrm{th}}}.$ This takes into account all directions with sufficiently strong radio emission.
After performing this integration, the expression for the angular aperture reduces to
\begin{equation}
\Delta \Omega_{CR}=
\cos^{-1}\left(\frac{\cos\theta_\mathrm{max}}{\sin i}\right)-\sin(\theta_\mathrm{max})\Bigl[\sin(\theta_\mathrm{max})
\cos(i) \phi_\mathrm{max} + \cos(\theta_\mathrm{max}) \sin(i) \sin(\phi_\mathrm{max})\Bigr],
\label{omega-cr}
\end{equation}
where $\cos(\theta_\mathrm{max})=1/n-D$ and  $\cos(\phi_\mathrm{max})=\frac{\cos(i)
\cos(\theta_\mathrm{max})}{\sin(i) \sin(\theta_\mathrm{max})}.$
The total aperture for cosmic rays is obtained
by substituting (\ref{omega-cr}) in (\ref{acr}) and integrating over the
polar angle $\theta_s$.

When the UHE primary is instead a neutrino, it can produce showers deep below the surface of the Moon and there will be considerable attenuation of the radio waves which travel distances longer than $\lambda_r$ below the surface. For the neutrino induced showers, the aperture is defined in the same way as for the CR, but the
angular aperture is now given \cite{gusev} by
\begin{equation}
\Delta\Omega_{\nu}(E,\theta_s)=\int \frac{\mathrm{d}z}{\lambda_{\nu}}
\int\Theta\bigl\{{\cal E}(E,\theta_s, \theta)\exp[-z/(2 \lambda_r \cos i)]-{\cal E}_{th}\bigr\} \times \exp[-L(z,\beta)/\lambda_{\nu}]
\times \mathrm{d}\alpha \mathrm{d}\cos\beta,
\label{omega-nu}
\end{equation}
where $L(z,\beta)$ is the distance the neutrino travels
inside the material to reach the interaction point at a
distance $z$ below the surface.
 In performing this integration we 
allow $z$ to go below the known depth of the regolith. 
The aperture can therefore pick up contributions 
from sufficiently strong signals coming from deep showers, 
especially for the lower frequencies. Numerically
we find for the worst case (when $\nu=150$ MHz), that imposing
a sharp cutoff at a depth of $20$~m would reduce the
aperture by nearly an order of magnitude, similarly to what was
discussed in \cite{Scholten:2005pp}.

It has been observed \cite{gusev}, that if the absorption length of radio waves $\lambda_r$
is much smaller than the neutrino mean free path $\lambda_{\nu}$ (which is indeed here the case) then
Eq.~(\ref{omega-nu}) reduces approximately to
\begin{equation}
\Delta\Omega_{\nu}(E,\theta_s)=\int\frac{2 \lambda_r
\cos(i)}{\lambda_{\nu}} \ln({\cal E}(E)/{\cal E}_{th}) \Theta[{\cal E}(E,\theta_s, \theta)-{\cal E}_\mathrm{th}]
\times \mathrm{d}\alpha \mathrm{d}\cos\beta.
\label{omega-nu1}
\end{equation}
As for the cosmic rays, the total aperture is obtained
by substituting (\ref{omega-nu}) into (\ref{acr}) and integrating over
the polar angle $\theta_s$.

To estimate the sensitivity of GMRT to cosmic ray and neutrino events we have evaluated the angular apertures by employing this technique and performing numerical integrations for the different parameters given in Table \ref{tab1}. In the next section we will discuss these results further in the context of prospective flux limits.

\section{Limits on the flux of cosmic rays and neutrinos}
\begin{figure}[t]
\hbox{\hspace{0cm}
\hbox{\includegraphics[scale=1.3]{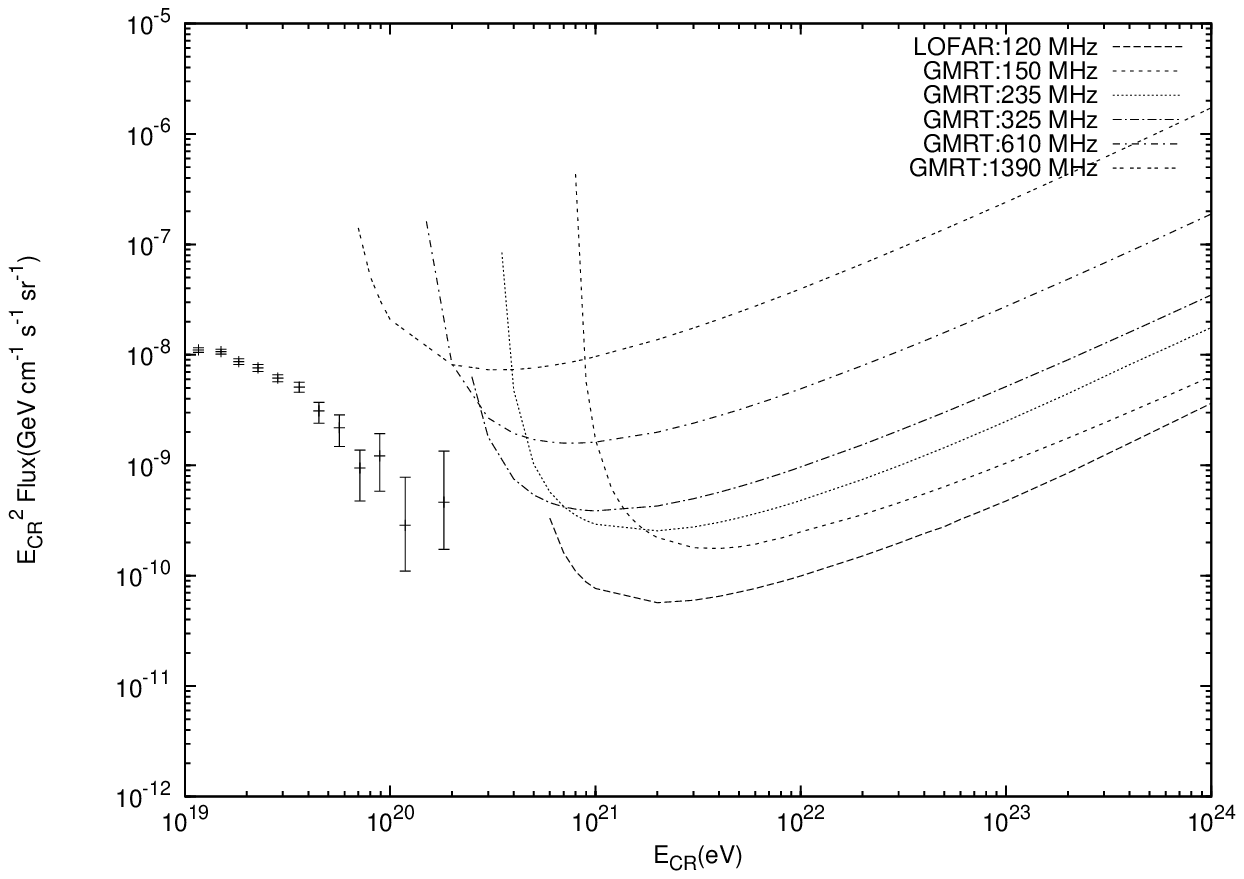}}}
\caption{Model independent limits on UHECR flux at different
frequencies for 100 hours of observation time with GMRT. Auger data points reproduced from \cite{SemikozAuger} on the CR flux are shown for comparison.}
\label{crlimit100}
\end{figure}

\begin{figure}[t]
\hbox{\hspace{0cm}
\hbox{\includegraphics[scale=1.3]{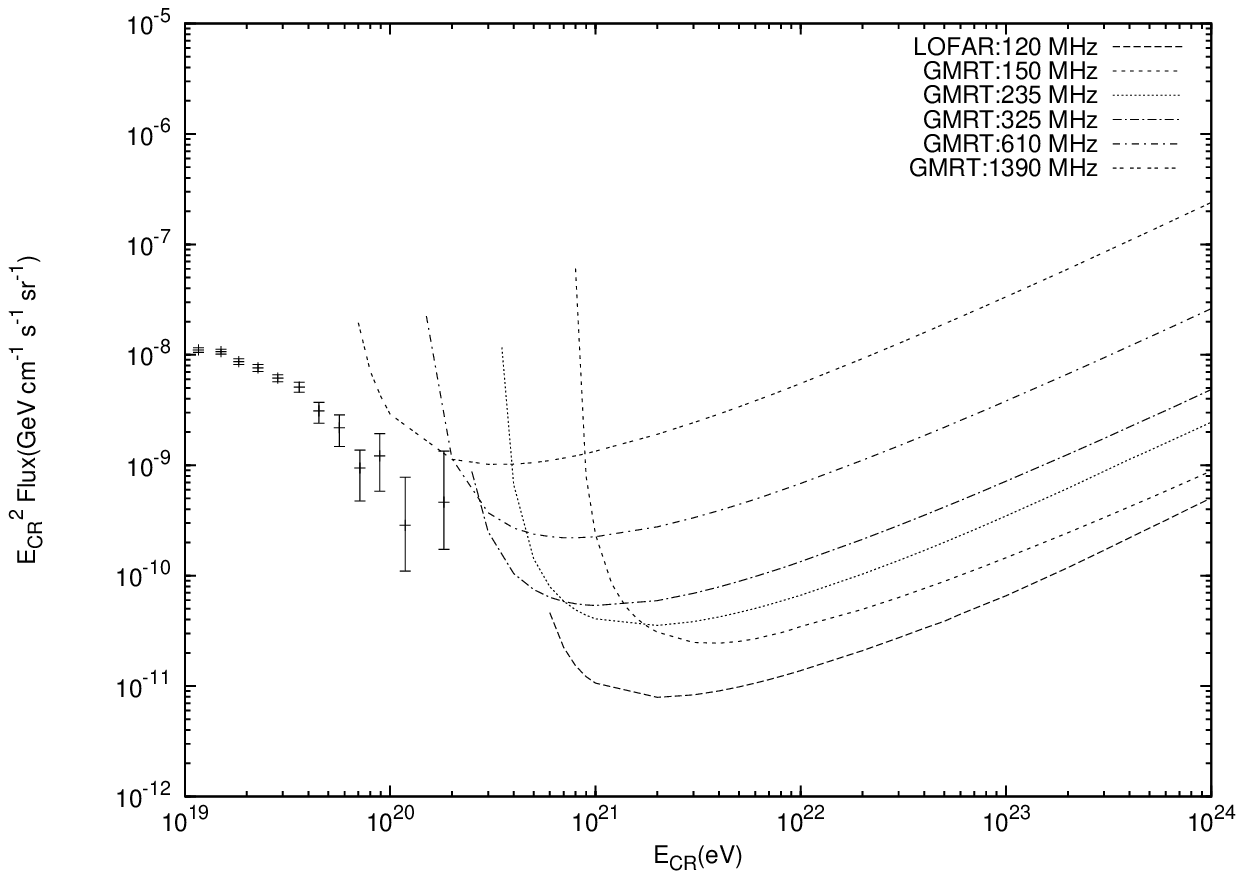}}}
\caption{Limits on UHECR flux at different
frequencies for 30 days of observation time with GMRT. Auger data as before.}
\label{crlimit30}
\end{figure}

Should no events be observed at GMRT during observation over a time $T$, an upper limit can be established on sufficiently smooth UHECR and neutrino fluxes at the Moon. The conventional model-independent limit \cite{FORTE} is given by
\begin{equation}
E_{i}^2 \frac{\mathrm{d}\Phi_{i}}{\mathrm{d}E_{\nu}} \leq s_\mathrm{up} \frac{E_{i}}{A_{i}(E_s=y_i E_{i})T},
\end{equation}
where still $i=\left\{\nu,\mathrm{CR}\right\}$, $y_{CR}=1$ and $y_\nu=0.25$. The Poisson factor $s_\mathrm{up}=2.3$ for a limit at $90 \%$ confidence level. The limits on the flux of UHECR that could be established for 100 hours and 30 days of observation time at GMRT are shown in Fig.~\ref{crlimit100} and Fig.~\ref{crlimit30} respectively. We also show the results of carrying through our calculations for the LOFAR parameters given in \cite{Scholten:2005pp}. When compared to the LOFAR Monte Carlo simulation results, the two methods of calculation agree within a factor two over the full energy range. This is an acceptable discrepancy in level with known uncertainties from {\it{e.g.}} the regolith depth. A residual difference of $\sim 30\%$ is also expected since we used the Gaussian approximation. The limit we obtain for the LOFAR parameters is less stringent than the one published, so in this respect our result constitutes a conservative estimate.
>From comparing the limits for different frequencies, it can be seen that low frequency observations give more stringent limits on the flux at the expense of a higher threshold. This is due to the well-known increase in the aperture \cite{Scholten:2005pp} from radiation spreading at lower frequencies.
The figures also show the Auger data \cite{SemikozAuger} on UHE cosmic rays. Although the extrapolation to higher energies is highly uncertain, the GMRT would most probably be sensitive to a post-GZK proton flux.

\begin{figure}[t]
\hbox{\hspace{0cm}
\hbox{\includegraphics[scale=1.3]{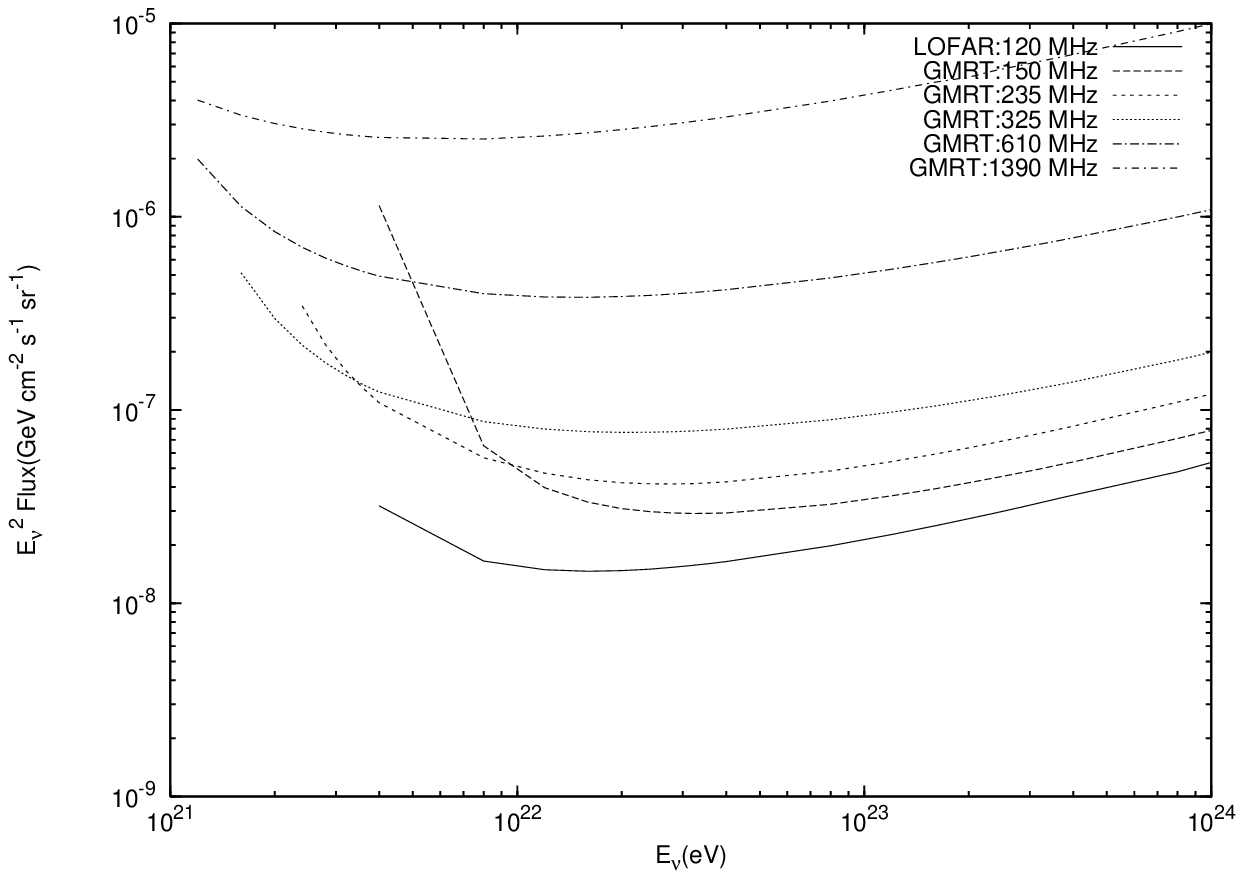}}}
\caption{Model independent limits on UHE neutrino flux
at different frequencies for 100 hours of observation time with GMRT. For comparison we show also a limit calculated for the same observation time with the LOFAR parameters of \cite{Scholten:2005pp}.}
\label{nlimit100}
\end{figure}

\begin{figure}[t]
\hbox{\hspace{0cm}
\hbox{\includegraphics[scale=1.3]{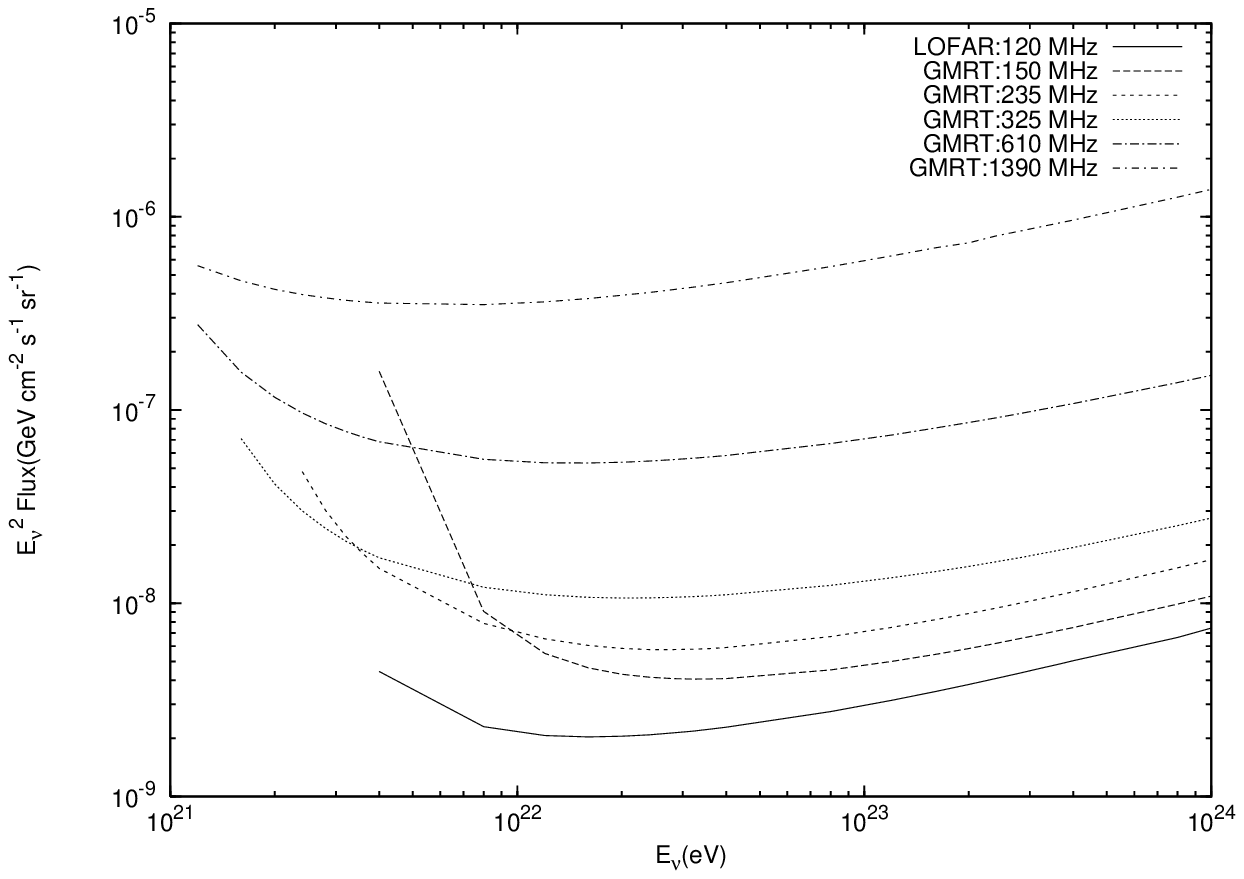}}}
\caption{Limits on UHE neutrino flux at different frequencies for 30 days of observation time with GMRT. LOFAR calculation with the same parameters as before but also for $T=30$ days.}
\label{nlimit30}
\end{figure}

Similarly for the UHE neutrinos, prospective limits on their flux for $T=100$ hours and $T=30$ days are shown in Figs.~\ref{nlimit100} and \ref{nlimit30}. Also here we show a calculation for the LOFAR parameters, again in quantitative agreement with previous results. Since many radio experiments exist for UHE neutrino detection, we have compiled a comparison in Fig.~\ref{fluxgmrt}. This figure contains, in addition to the GMRT results for $\nu=150$ MHz with two different observation times, the already existing limits from RICE \cite{rice}, GLUE \cite{glue}, FORTE \cite{FORTE} and ANITA-lite \cite{anita}. Also we have indicated the prospective future limits that has been calculated for ANITA \cite{anita}, LOFAR \cite{Scholten:2005pp} or LORD \cite{gusev}.

\section{Summary and Conclusions}
\begin{figure}[t]
\hbox{\hspace{0cm}
\hbox{\includegraphics[scale=1.3]{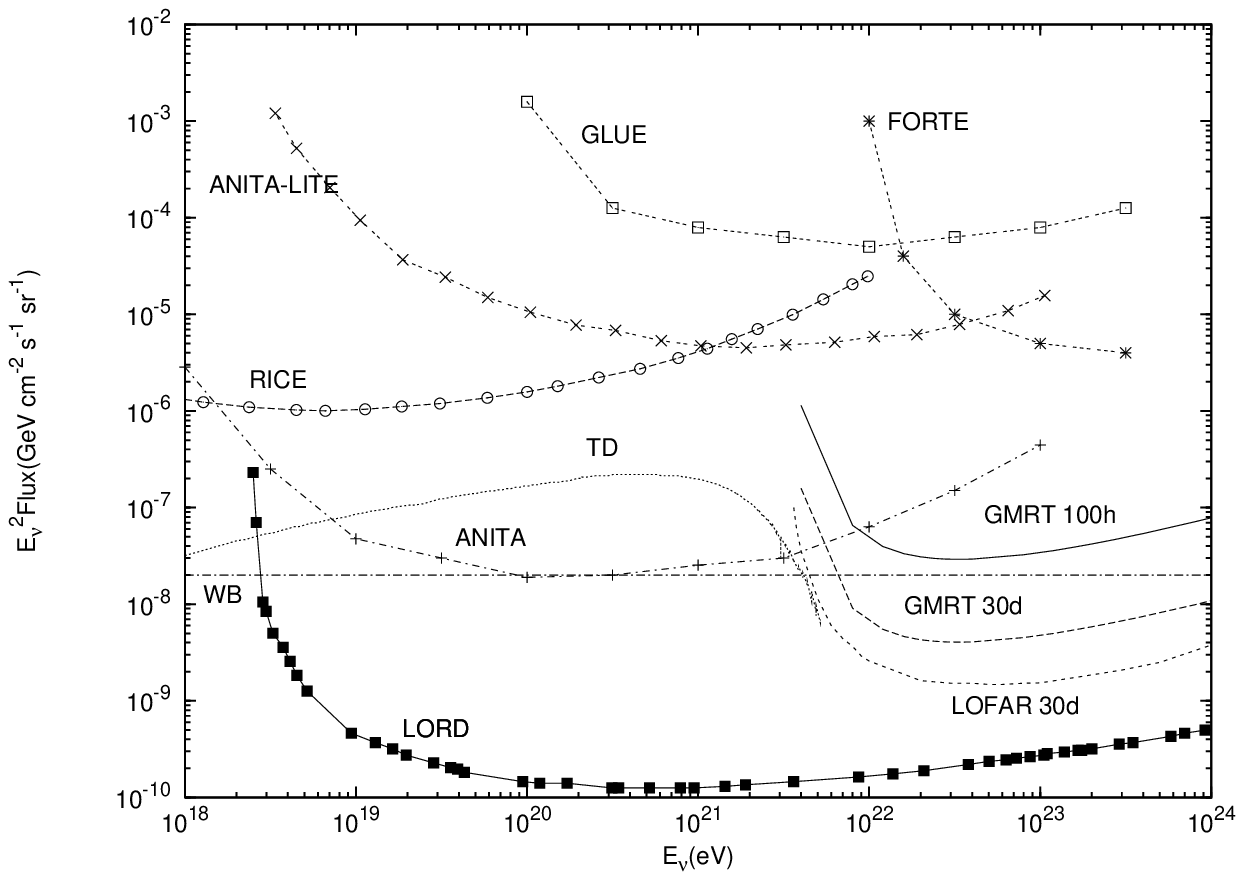}}}
\caption{Prospective flux limits on UHE neutrinos from 
GMRT shown for effective exposure times of $100$ hours and $30$ days. 
The current best limits from radio experiments 
ANITA-lite \cite{anita}, GLUE \cite{glue}, FORTE \cite{FORTE}, and 
RICE \cite{rice} are shown. For comparison the expected limits 
from future experiments ANITA \cite{anita}, LOFAR \cite{Scholten:2005pp} 
and LORD \cite{gusev} are also included. The WB line indicates 
the theoretical upper limit of Waxman-Bahcall \cite{Waxman:1998yy} 
on the cosmogenic neutrino flux. TD refers to the Topologi
cal Defect model with $M_X=2\times10^{22}$ eV described in \cite{Semikoz&Sigl}.}
\label{fluxgmrt}
\end{figure}

We have calculated the potential for GMRT to detect 
UHE cosmic rays and neutrinos through their radio 
wave emission produced when showering in the lunar 
regolith. Our results indicate that GMRT could be 
competitive to future experiments in the $E\gtrsim 10^{20}$ eV range. 
If one assumes that the CR spectrum continues beyond the 
GZK limit with unchanged energy dependence, observation of 
these particles with the GMRT should indeed be possible.
For UHE neutrinos there exists a theoretical upper bound 
on the flux from cosmogenic sources, as given by 
Waxman and Bahcall \cite{Waxman:1998yy}: $E_\nu^2 \Phi_\nu < 2 \times
10^{-8}$\, GeV\,cm$^{-2}$\, s$^{-1}$\, sr$^{-1}$. The GMRT could, 
using a mere 30 days of observation time, probe fluxes a factor five smaller.
As a benchmark scenario, we indicate in Fig.~\ref{fluxgmrt} the predicted flux from one TD model \cite{Semikoz&Sigl} where the mass scale resides around $10^{22}$ eV. 

It is notable that the GMRT (for the low frequencies) has 
only a somewhat higher threshold, and slightly worse sensitivity 
for this type of experiment, than will the LOFAR telescope. This clearly points to the interesting potential of utilizing the GMRT for UHE particle searches in the near future. The results we have presented here obviously depend on to what extent the experimental realization of these measurements are possible at the GMRT facility. We therefore foresee a future analysis taking more thoroughly into account the requirements on the technical infrastructure and the experimental techniques for signal identification and background discrimination.

\acknowledgments{The authors thank Gunnar Ingelman for reading the manuscript, and for providing insightful comments and suggestions. We also thank Olaf Scholten for useful communication.}



\bibliographystyle{sort}

\end{document}